 \documentclass[final,5p,times,twocolumn]{elsarticle}
\usepackage{amssymb}
\usepackage{lipsum}
\usepackage{graphicx}% Include figure files
\usepackage{dcolumn}% Align table columns on decimal point
\usepackage{bm}% bold math
\usepackage{siunitx}
\usepackage{amsmath}
\usepackage{subcaption}
\usepackage{comment}
\usepackage{multirow}
\newcommand{\sNN}{$\sqrt{s_{\rm NN}}$}
\newcommand{\pt}{$p_{\rm T}$}

\journal{Physics Letters B}

\begin{document}

\title{Beam energy dependence of net-hyperon yield and its implication on baryon transport mechanism}% Force line breaks with \\

\author[Kent,BNL]{Chun Yuen Tsang}
\author[BNL]{Rongrong Ma}
\author[BNL]{Prithwish Tribedy}
\author[Kent,BNL]{Zhangbu Xu}

\affiliation[Kent]{organization={Kent State University},%Department and Organization
            addressline={Department of Physics}, 
            city={Kent},
            postcode={44242}, 
            state={OH},
            country={USA}}

\affiliation[BNL]{organization={Brookhaven National Laboratory},%Department and Organization
            addressline={Physics Department}, 
            city={Upton},
            postcode={11973}, 
            state={NY},
            country={USA}}

\date{\today}% It is always \today, today,

\begin{abstract}

In the constituent quark model, each quark inside a baryon carries 1/3 unit of the baryon number. An alternative picture exists where the center of a Y-shaped topology of gluon fields, called the baryon junction, carries a unit baryon number. Studying baryon transport over a large rapidity gap ($\delta y$) in nuclear collisions provides a possible tool to distinguish these two pictures. A recent analysis of global data on net-proton yield at mid-rapidity in Au+Au collisions showed an exponential dependence on $\delta y$ and the exponential slope does not vary with event centrality, favoring the baryon junction picture. Since junctions are flavor blind, hyperons -- baryons containing valence strange quarks -- are expected to exhibit a similar behavior as the proton. This study aims to test this prediction by analyzing hyperon yields in Au+Au collisions at various energies. We observe that net-hyperon yields, after correcting for the strangeness production suppression, adhere to the expected exponential form. The extracted slope parameters for net-$\Lambda$, net-$\Xi$ and net-$\Omega$ are consistent with each other and with those of net-proton within uncertainties, and exhibit no centrality dependence, further substantiating the baryon junction picture. Various implementations of the \texttt{PYTHIA} event generator, primarily based on valence quarks for baryon transport, are unable to simultaneously describe the slope parameters for all baryons. 
\end{abstract}

\maketitle

%\tableofcontents

\section{Introduction}

The baryon number ($B$) is a conserved quantity in quantum chromodynamics (QCD). While mesons, like pions and kaons, carry zero baryon number ($B=0$), baryons such as protons and neutrons carry one unit ($B=1$). Since baryons are composed of three valence quarks, the conventional quark model assigns 1/3 unit of the baryon number to each quark. However, an alternative theory proposes that the entire unit of the baryon number is carried by the baryon junction, a non-perturbative Y-shaped topology of gluons connected to all three valence quarks~\cite{Rossi:1977cy, Artru:1990wq}. This junction is the only gauge-invariant state vector for a baryon made of three valance quarks and gluons~\cite{Kharzeev:1996sq}, and its existence receives support from lattice QCD calculations~\cite{Suganuma:2004zx, Takahashi:2000te}. Neither theory has been verified experimentally since many physical processes evolve similarly regardless whether the junction or valence quarks carry the baryon number.  However, recent proposals have outlined experimental approaches to distinguish these scenarios in high-energy hadronic and photon-induced collisions~\cite{lewis2023search,Frenklakh:2023pwy,Lv:2023fxk,Dong:2023zbu, Frenklakh:2024mgu,Magdy:2024dpm,Pihan:2024lxw,STAR:2024lvy}. 

One crucial difference between valence quark and baryon junction pictures is how efficient it is to transport them from beam rapidity ($y_{\text{beam}}$) to mid-rapidity ($y\sim0$). It has been argued that it is much easier to transport baryon junctions over a large rapidity gap than valence quarks in high-energy heavy-ion collisions, since junctions consist of low-momentum gluons and thus have sufficient time to interact and be stopped by the other colliding nucleus~\cite{Kharzeev:1996sq}. In contrast, valence quarks carry most of the momentum of an incoming nucleon and tend to pass through the other colliding nucleus unhindered. Therefore, baryon transport in heavy-ion collisions can be utilized to potentially distinguish the two pictures. The magnitude of the baryon transport can be quantified by measuring net-baryon yield, $i.e.$, baryon yield minus anti-baryon yield. This is because the baryon number is strictly conversed, so any excess of baryons over anti-baryons must originate from colliding nuclei.

For the baryon junction picture, the Regge theory predicts an exponential decrease in net-baryon yield against the rapidity gap ($\delta y$) over which baryons are transported, $i.e.$, $(dN/d\delta y)_\text{net-baryon} \propto e^{-\alpha_B\delta y}$. Here, $\delta y = y_\text{beam} - y$, where $y$ is the rapidity at which the net-baryon yield is measured, and the slope parameter $\alpha_B \sim 0.42 - 1$~\cite{Kharzeev:1996sq}.
%On the other hand, the \texttt{PYTHIA} 6 event generator, which assigns the baryon number to valence quarks, predicts a significantly stronger dependence on $\delta y$ with $\alpha_B \sim 2.5$ in $\gamma$+$p$ events \cite{lewis2023search}. In $p$+$p$ collisions, event generators such as \texttt{PYTHIA} and \texttt{HERWIG}, which are based on valence quarks, have been extensively tuned to align with observed baryon production. The tuning varies, leading to different predicted slope parameters~\cite{lewis2023search}. 

%Experimentally, the net-baryon yield can be approximated with the net-proton yield since (anti-)protons make up a large portion of (anti-)baryons. 
A recent analysis of net-proton (net-$p$) yields at mid-rapidity in Au+Au collisions at various beam energies from the Beam Energy Scan Phase-I (BES-I) program~\cite{STAR:2008med, STAR:2017sal, Chen:2024zwk} conducted at Relativistic Heavy Ion Collider (RHIC) revealed an exponential behavior as a function of $\delta y$~\cite{lewis2023search}. The net-$p$ yields include both the primordial production and those from weak decays of hyperons, $i.e.$, baryons containing at least one valence strange quark, and therefore they can be regarded as good proxies of the total net-baryon yields. The extracted exponential slope for the 0-80\% collision centrality is $\alpha_B = 0.64\pm0.05$~\cite{lewis2023search}. Here, the term ``centrality" is used to describe the geometrical overlap between the two colliding nuclei~\cite{Miller:2007ri}, with 0\% (100\%) corresponding to most central (peripheral) collisions with smallest (largest) impact parameter and most (least) number of multiple scatterings. Such a slope is consistent with the baryon junction prediction. It has been shown in Ref.~\cite{lewis2023search} that the measured $\alpha_B$ for net-$p$ is also consistent with specific tunes of the \texttt{PYTHIA} event generator for proton+proton ($p$+$p$) collisions, which assigns baryon number to valence quarks. However, the extracted $\alpha_B$ value in data shows little dependence on centrality~\cite{lewis2023search}, suggesting that the multiple scattering effect does not impact the baryon transport. This is inconsistent with the valence quark picture as one would expect the stronger increase of multiple scattering with increasing beam energy in central collisions should result in larger valence quark transport, thus a smaller slope, than that in peripheral collisions. 
%Reference~\cite{lewis2023search} posits these observations as evidence for the baryon junction being the baryon number carrier. 

Another unique feature of the baryon junction is that it is flavor blind~\cite{lewis2023search}, meaning once a junction is pulled out of the incoming nucleus, baryons of any flavor can emerge around it modulo different energies needed to produce quarks of different masses. Consequently, one would expect the transport behavior of hyperons to resemble that of inclusive baryons. In this manuscript, we compile published data on hyperon production at mid-rapidity in heavy-ion collisions of various beam energies. By analyzing the transport behavior of hyperons, we aim to test the flavor independence of the baryon junction mechanism. Specifically, we will inspect if the dependence of net-$\Lambda$, net-$\Xi$ and net-$\Omega$ yields on $\delta y$ are consistent with each other and with that of net-$p$ by comparing slope parameters ($\alpha_B$) as a function of centrality. Here, the dependence on $\delta y$ originates from variations in $y_\text{beam}$ for different collision energies while the yields are always measured at mid-rapidity ($y\sim0$). Furthermore, predictions of $\alpha_B$ for net-hyperons from various versions and tunes of the \texttt{PYTHIA} event generator~\cite{Sjostrand:2006za, Skands:2010tm, Strand:2008ab}, including those that can describe the slope parameter for net-$p$, will be compared to our results.

\section{Beam energy dependence of net-hyperon yields}

Using published data from RHIC BES-I program, we compute net-$\Lambda$, net-$\Xi$ and net-$\Omega$ yields, $(dN/dy)_{\Lambda - \bar{\Lambda}}/\langle N_\text{part}\rangle$, $(dN/dy)_{\Xi - \bar{\Xi}}/\langle N_\text{part}\rangle$, and $(dN/dy)_{\Omega - \bar{\Omega}}/\langle N_\text{part}\rangle$, within the rapidity range of $|y| < 0.5$ in Au+Au collisions at center-of-mass energies (\sNN) of 7.7, 11.5, 19.6, 27, 39, 62.4 and $\SI{200}{GeV}$~\cite{star:2007sp, star:2012se, star:2019sh, STAR:2010yyv}. Here $N_\text{part}$ is the average number of participating nucleons in a given centrality class of Au+Au collisions. It is used for normalization such that measurements in nearby centrality classes can be combined in extracting the slope parameter when needed. For instance, the net-$\Lambda$ yield is measured for 0-5\% centrality class at 200 GeV, but for 0-10\% at other energies. For studying the $\delta y$ dependence of $(dN/dy)_{\Lambda - \bar{\Lambda}}/\langle N_\text{part}\rangle$, we treat these two data points as part of the same group. 

%Unlike up and down quarks, which are already present in the colliding nucleus, strange quarks must be pair-produced. Near the production threshold, $dN/dy_\text{net-hyperon}$ will naturally decrease with decreasing energy because strange quarks production are reduced, leading to a reduction in (anti-)hyperon yields. However, the expected exponential dependence from the baryon junction picture implies a monotonically increasing trend in net-hyperon yields as beam energy decreases (assuming that $\alpha_B > 0$). As a result, even without considering any experimental data, one expects the exponential dependence on $\delta y$ to break down due to the difficulties in pair production. It is imperative to account for the effects associated with the strange quark production to isolate the effect of baryon transport.

\begin{figure}[h!]
\centering
    \includegraphics[width=0.85\linewidth]{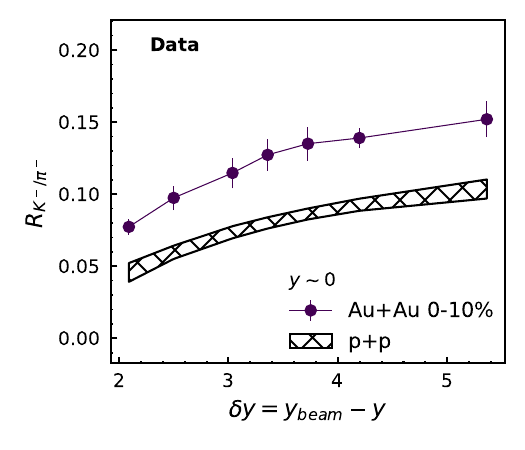}
    \caption{$K^-/\pi^-$ yield ratio at mid-rapidity ($R_{K^-/\pi^-}$) as a function of $\delta y$ in 0-10\% central Au+Au~\cite{STAR:2017sal} and $p$+$p$ collisions. Vertical error bars around data points represent combined statistical and systematic uncertainties. The hatched band indicates the parametrization uncertainties for $p$+$p$ collisions. }
    \label{fig:dataKpi}
\end{figure}

Compared to up and down quarks that make up of protons, it is more difficult to produce strange quarks for forming hyperons due to their larger mass. Such an effect is also influenced by the beam energy, and therefore needs to be accounted for in comparing the transport properties of baryons with varying number of valence strange quarks. We gauge the difficulty of strange quark production using the $K^-(s\bar{u})/\pi^-(d\bar{u})$ yield ratio, denoted as $R_{K^{-}/\pi^{-}}$, in 0-10\% central Au+Au collisions. Such a ratio is conventionally considered representing the ratio of strangeness to entropy in the system~\cite{NA49:2002ed}. The $K^+/\pi^+$ ratio is not used since the $K^+$ yield is enhanced by the associated production ($p + N \rightarrow \Lambda + K^+ + N$).
%\begin{equation}
%p + N \rightarrow \Lambda + K^+ + N.
%\end{equation}
At low beam energies, this contribution even dominates over the pair production for $K^+$~\cite{Khan:2022}. 
%It has been observed that the $K^+/\pi^+$ ratio peaks at $\sqrt{s_{NN}}\approx\SI{10}{GeV}$ across various heavy-ion collision systems~\cite{STAR:2017sal} and exhibits non-monotonic behavior as a function of beam energies. 
Given our focus on strange quarks that are pair-produced as the junction comes to a halt, the $K^-/\pi^-$ ratio serves as a better proxy for strangeness production suppression than the $K^+/\pi^+$ ratio. It is worth noting that this approach implicitly assumes that the hot medium created in Au+Au collisions affects the strange mesons and baryons the same way. $R_{K^-/\pi^-}$ within the rapidity range of $|y| < 0.1$ is extracted from published RHIC BES-I data~\cite{STAR:2017sal}, and shown as a function of $\delta y$ in Fig.~\ref{fig:dataKpi}. The ratio is seen to increase monotonically with increasing beam energy, consistent with our expectation that more strange quarks are produced at higher beam energies. We also investigate using $R_{K^-/\pi^-}$ from $p$+$p$ collisions, as an extreme case, for gauging the strangeness production suppression. Parametrization of $R_{K^-/\pi^-}$ at mid-rapidity as a function of collision energy in $p$+$p$ collisions is detailed in Appendix, and included in Fig.~\ref{fig:dataKpi} for comparison. The monotonic increase as a function of $\delta y$ is also observed for $p$+$p$ collisions, but the ratio is consistently lower than that from Au+Au collisions. Furthermore, the $\delta y$ dependence in $p$+$p$ appears to be weaker than that in Au+Au collisions, which could be due to the presence of hot medium effects in Au+Au collisions. 

\begin{figure*}[h!]
\centering
\begin{subfigure}[b]{0.32\linewidth}
    \includegraphics[width=\linewidth]{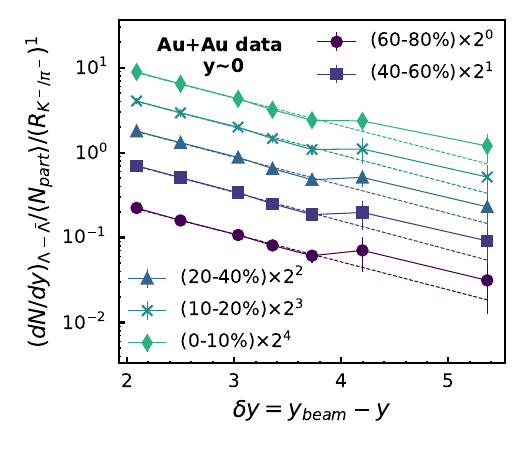}
    \caption{}
    \label{fig:rawNetd}
\end{subfigure}
\begin{subfigure}[b]{0.32\linewidth}
    \includegraphics[width=\linewidth]{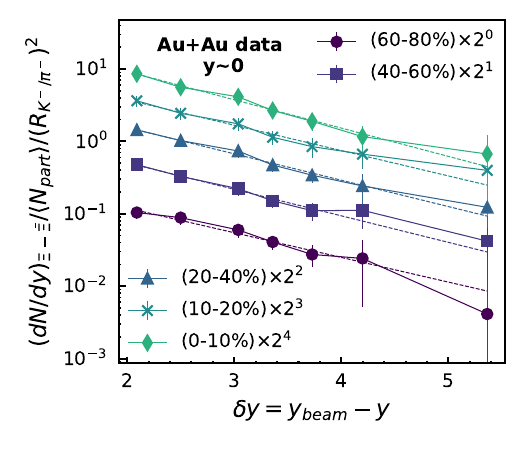}
    \caption{}
    \label{fig:rawNete}
\end{subfigure}
\begin{subfigure}[b]{0.32\linewidth}
    \includegraphics[width=\linewidth]{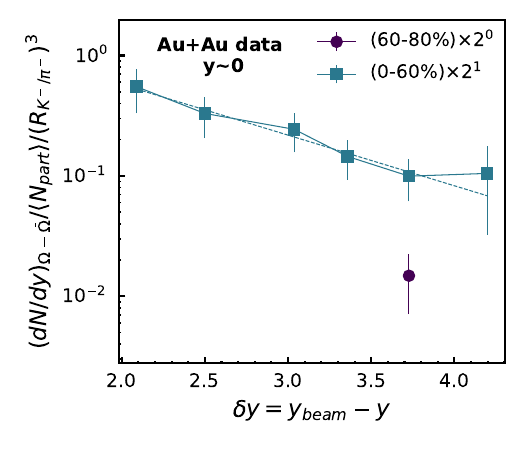}
    \caption{}
    \label{fig:rawNetf}
\end{subfigure}
\caption{Net-$\Lambda$ (a), net-$\Xi$ (b) and net-$\Omega$ (c) yields within $|y| < 0.5$ as a function of rapidity gap in Au+Au collisions. These yields are scaled by $(R_{K^-/\pi^-})^n$ in 0-10\% central Au+Au collisions, where $n$ is the number of valence strange quarks in the hyperon. Vertical error bars around data points represent combined statistical and systematic uncertainties. Data at each centrality are fitted with an exponential function, shown as a dotted line. }
\label{fig:corNet}
\end{figure*}

The correction for the strangeness production suppression is performed by dividing the net-hyperon yields by $(R_{K^-/\pi^-})^n$, where $n$ represents the number of valence strange quarks in the hyperon, $i.e.$, $n=1$ for net-$\Lambda$, $n=2$ for net-$\Xi$ and $n=3$ for net-$\Omega$.
%Using published hyperon yields from RHIC BES-I, we compute net-hyperon yields, $dN/dy|_{\Lambda - \bar{\Lambda}}/\langle N_\text{part}\rangle/(K^-/\pi^-)$, $dN/dy|_{\Xi - \bar{\Xi}}/\langle N_\text{part}\rangle/(K^-/\pi^-)^2$, and $dN/dy|_{\Omega - \bar{\Omega}}/\langle N_\text{part}\rangle/(K^-/\pi^-)^3$, with hyperon yields measured within the rapidity range of $|y| < 0.5$ in Au+Au collisions at center-of-mass energies (\sNN) of 7.7, 11.5, 19.6, 27, 39, 62.4 and $\SI{200}{GeV}$~\cite{star:2007sp, star:2012se, star:2019sh, STAR:2010yyv}. $N_\text{part}$ is the average number of participating nucleons in a given centrality class of Au+Au collisions. It is used for normalization such that measurements in nearby centrality classes can be combined in extracting the slope parameter when needed. For instance, the net-$\Lambda$ yield is measured for 0-5\% centrality class at 200 GeV, but for 0-10\% at other BES-I energies. For studying the energy or beam rapidity dependence of $dN/dy|_{\Lambda - \bar{\Lambda}}/\langle N_\text{part}\rangle$, we treat these two data points as part of the same group.
The normalized net-hyperon yields using $R_{K^-/\pi^-}$ in 0-10\% central Au+Au collisions are plotted as a function of $\delta y = y_\text{beam} - y$ in Figs.~\ref{fig:rawNetd},~\ref{fig:rawNete} and~\ref{fig:rawNetf}, respectively, where $y$ is set to 0 since the hyperon yields are measured at mid-rapidity. Systematic uncertainties in the measured hyperon and anti-hyperon yields are assumed to be uncorrelated, and the combined statistical and systematic uncertainties for net-hyperon yields are shown as vertical bars around data points in the figure. Since net-hyperon yields are nearly flat within $|y| < 0.5$ for beam energies relevant for this work~\cite{BRAHMS:2005cm, NA49:2002ed, E802:1996pp}, the difference in the rapidity window for net-hyperon yields and $R_{K^-/\pi^-}$ ($|y| < 0.5$ vs. $|y| < 0.1$) should have negligible impact. 
%The Kaon to pion ratios were published with finer centrality bins, so we take an average of nearby bins to reduce the data to the coarser centrality bins used in hyperon data. Uncertainties were taken as the average uncertainties of the summed bin, rather than propagating them as if they were independent variables due to the highly correlated nature of particle ratios in neighboring centrality bins. 

\begin{figure}[h!]
\centering
    \includegraphics[width=0.85\linewidth]{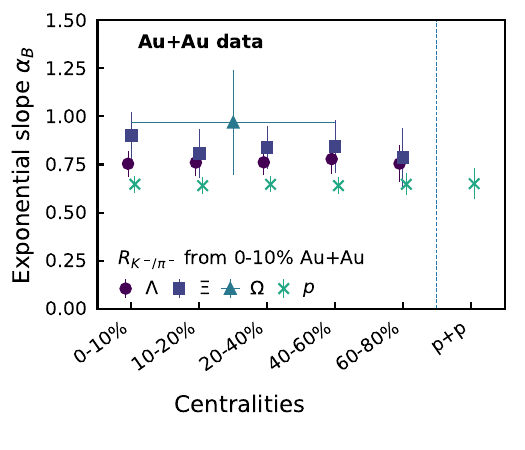}
    \caption{Comparison of slope parameters for net-$\Lambda$, net-$\Xi$, net-$\Omega$ and net-$p$ as a function of centrality from Au+Au collisions at \sNN\ = 7.7 - 200 GeV. The value for net-$\Omega$ is from 0-60\% centrality interval. The $\alpha_{B}$ value for inclusive net-$p$ in $p$+$p$ collisions is also shown for comparison.}
    \label{fig:AllData}
\end{figure}

The dotted lines in Fig.~\ref{fig:corNet} represent exponential fits to scaled net-hyperon yields: 
\begin{equation}
 (dN/dy)_\text{net-hyperon}/\langle N_\text{part}\rangle/(R_{K^-/\pi^-})^n \propto e^{-\alpha_B\delta y}.
 \label{eq:exp}
 \end{equation} 
The fitted slope parameters for all centrality classes and particle species are summarized in Fig.~\ref{fig:AllData}.
%This figure shows that not only do the data points closely resemble an exponential function, but the fitted $\alpha_B$ values in central collision between different hyperon species are consistent with each other within uncertainties. 
%The $\alpha_B$ values for different hyperons are: net-$\Lambda: 0.75 - 0.97$, net-$\Xi: 0.82 - 1.15$, and net-$\Omega: 1.01 \pm 0.12$. The ranges for net-$\Lambda$ and net-$\Xi$ represent the minimum to maximum fitted $\alpha_B$ values across different centralities, while the uncertainty in net-$\Omega$ represents the fitting uncertainty in 0-60\% centrality. 
%The similarity of $\alpha_B$ values across all three hyperon species suggests a flavor-blind transport mechanism. %Although $\alpha_B$ values for different hyperon species are still consistent within uncertainty, 
$\alpha_B$ values for net-hyperons are consistent with those of inclusive net-$p$ within uncertainties, providing strong evidence of flavor-blind baryon transport as predicted by the baryon junction picture. Furthermore, no centrality dependence is seen for net-$\Lambda$ and net-$\Xi$ slope parameters, similar to the case of net-$p$. As aforementioned, such a centrality independence is incompatible with valence quark transport. Also shown in Fig. \ref{fig:corNet} is $\alpha_{B}$ for inclusive net-$p$ in $p$+$p$ collisions, which is extracted based on data from the Intersecting Storage Rings (ISR) and RHIC~\cite{Banner:1972eja,British-Scandinavian:1974lzd,Antinucci:1972ib,STAR:2008med}. Details of the net-$p$ yield calculation and $\alpha_{B}$ extraction can be found in the Appendix. The $\alpha_{B}$ value in $p$+$p$ collisions is well in line with those in Au+Au collisions, confirming the independence of $\alpha_{B}$ on multiple scatterings. 

\begin{figure}[h!]
\centering
    \includegraphics[width=0.85\linewidth]{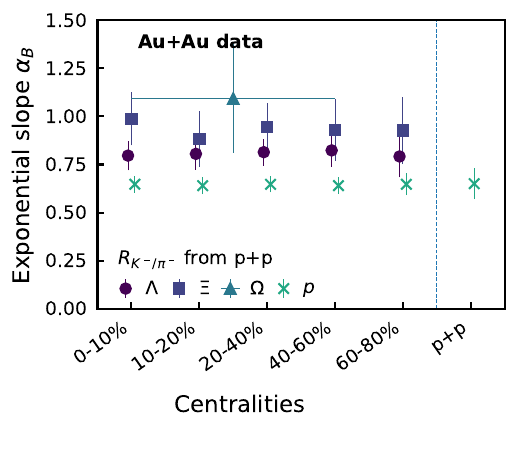}
    \caption{Same as Fig.~\ref{fig:AllData}, but fitted to net-hyperon yields corrected for strangeness production suppression using $R_{K^-/\pi^-}$ from $p$+$p$ collisions.}
    \label{fig:AllDatappPhiK}
\end{figure}

 The limiting case, in which $R_{K^{-}/\pi^{-}}$ from $p$+$p$ collisions are used to account for strange quark production suppression, is shown in Fig.~\ref{fig:AllDatappPhiK}. In this case, possible hot medium effects on strange quark production in heavy-ion collisions are ignored. Despite of the caveat, we see that the $\alpha_{B}$ values for different species of hyperons are still consistent with each other and with those of net-$p$. We also observe that separations in central values of $\alpha_{B}$ among different particle species are slightly larger than those when using $R_{K^{-}/\pi^{-}}$ from 0-10\% Au+Au collisions as shown in Fig. \ref{fig:AllData}.

\begin{figure*}[h!]
\centering
\begin{subfigure}[b]{0.32\linewidth}
    \includegraphics[width=\linewidth]{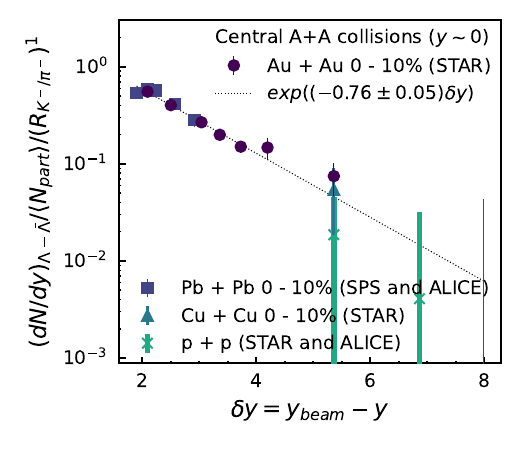}
    \caption{}
\end{subfigure}
\begin{subfigure}[b]{0.32\linewidth}
    \includegraphics[width=\linewidth]{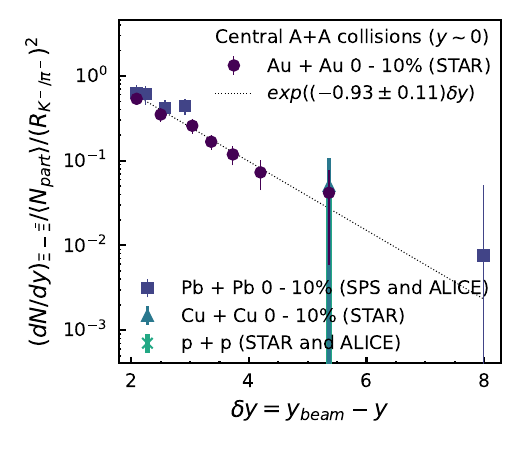}
    \caption{}
\end{subfigure}
\begin{subfigure}[b]{0.32\linewidth}
    \includegraphics[width=\linewidth]{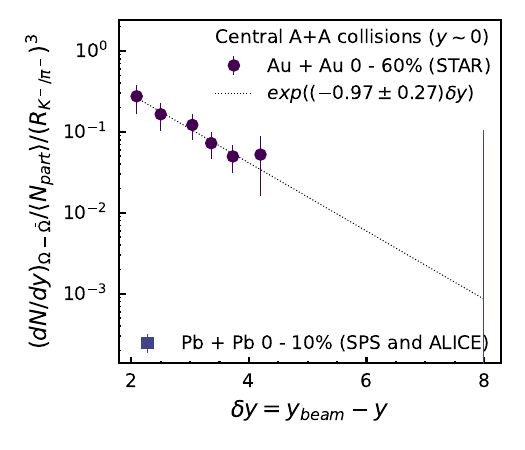}
    \caption{}
\end{subfigure}
\caption{Same as Fig.~\ref{fig:corNet}, but only for the most central collisions and with SPS and ALICE data on Pb+Pb collisions~\cite{NA49:2008ed, NA57:2006eo,Schuchmann2016,Abelev2014ms,Abelev2013ks}, ALICE and STAR data on $p$+$p$ collisions~\cite{STAR:2006nmo,ALICE:2010vtz}, and STAR data on Cu+Cu collisions~\cite{star:2012se} included. The marker representing $p$+$p$ collisions in panel (b) is not visible because its central value is negative. }
\label{fig:NetCompare}
\end{figure*}

Besides those from RHIC BES-I program (Fig.~\ref{fig:corNet}), hyperon yields are also available for Pb+Pb collisions of various beam energies as measured by NA49 Collaboration at CERN Super Proton Synchrotron (SPS) ~\cite{NA49:2008ed, NA57:2006eo} and by ALICE at the Large Hadron Collider (LHC)~\cite{Schuchmann2016,Abelev2014ms,Abelev2013ks}, for 200 GeV Cu+Cu collisions by STAR at RHIC ~\cite{star:2012se}, and for 200 GeV and 900 GeV $p$+$p$ collisions by STAR at RHIC~\cite{STAR:2006nmo} and ALICE at LHC~\cite{ALICE:2010vtz}, respectively. The world data on net-hyperon yields are shown in Fig.~\ref{fig:NetCompare} for $p$+$p$ and central heavy-ion collisions. $R_{K^-/\pi^-}$ for Pb+Pb collisions are only available at \SI{40}{AGeV}, \SI{80}{AGeV}, \SI{158}{AGeV}~\cite{Afanasiev2002ed} and \sNN\ = \SI{2.76}{TeV}~\cite{Abelev2013cd} within $|y| < 0.5$. For net-hyperon yields from Pb+Pb collisions at beam energies where $R_{K^-/\pi^-}$ are not available, we use $R_{K^-/\pi^-}$ value at the closest beam energy as the normalization factor. For Cu+Cu collisions, $R_{K^-/\pi^-}$ from 0-10\% 200 GeV Au+Au centrality class is used. For $p$+$p$ collisions, $R_{K^-/\pi^-}$ at 200 and 900 GeV are readily available from Refs.~\cite{star:2007sp, ALICE:2011gmo}. Exponential fits to the world data are performed as in Fig. \ref{fig:corNet}, and the resulting fit functions are shown as dotted lines in Fig. \ref{fig:NetCompare}. The fitted $\alpha_B$ values are consistent with those obtained with Au+Au data alone as shown in Fig. \ref{fig:AllData}, suggesting a universal transport mechanism in all the collision systems and energies explored here.

\section{Comparison between data and \texttt{PYTHIA}}

%Fragments yielding predictions for the $p$+$p$ system at various beam energies are generated from different versions of \texttt{PYTHIA} for comparison to experimental data.
In this section, we compare slope parameters in $p$+$p$ collisions predicted by the \texttt{PYTHIA} event generator, which employs valence quarks as the baryon number carriers, to those in Au+Au collisions. Such a comparison is justified given the near independence of the slope parameters on event multiplicity as illustrated in Figs. \ref{fig:AllData}. The versions and tunes tested include \texttt{PYTHIA} 6.4 default tune, \texttt{PYTHIA} 6.4 with Perugia version 0 (P0), \texttt{PYTHIA} 6.4 with Perugia version 2012 (P12)~\cite{Sjostrand:2006za, Skands:2010tm, Strand:2008ab}, \texttt{PYTHIA} 8.3 default tune and \texttt{PYTHIA} 8.3 with Color Reconnection (CR) Mode 2 beyond the leading-color approximation~\cite{Bierlich:2022pfr,Christiansen:2015yqa}. The Perugia tunes for \texttt{PYTHIA} 6.4 involve manual adjustments to match Tevatron minimum-bias and Drell-Yan data, as well as SPS minimum-bias data~\cite{Skands:2010tm}. While most of the \texttt{PYTHIA} tunes produce baryons mainly through the ``popcorn" mechanism, the version 8.3 with CR is of particular interest because it partially implements the baryon junction mechanism by allowing dynamic formation of baryon junction through color reconnection prior to hadronization, leading to enhanced baryon production at mid-rapidity. However, it still deviates from the real baryon junction picture by not including baryon junctions in the colliding protons and not simulating scatterings involving junctions~\cite{Christiansen:2015yqa}. Nevertheless, such a partial implementation could still provide insights into the influence of the baryon junction mechanism on baryon transport.

\begin{figure}[h!]
\centering
    \includegraphics[width=0.85\linewidth]{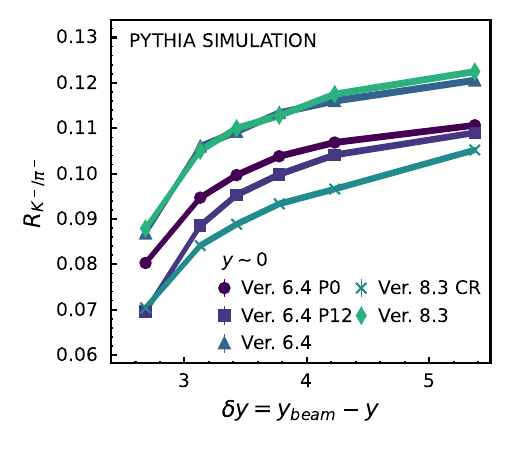}
\caption{$R_{K^-/\pi^-}$ within $|y| < 0.1$ as a function of $\delta y$ for different versions and tunes of the \texttt{PYTHIA} event generator. The width of the curves connecting the markers represents statistical uncertainties.}
    \label{fig:pythiaKpi}
\end{figure}

\begin{figure*}[h]
\centering
\begin{subfigure}[b]{0.32\linewidth}
    \includegraphics[width=\linewidth]{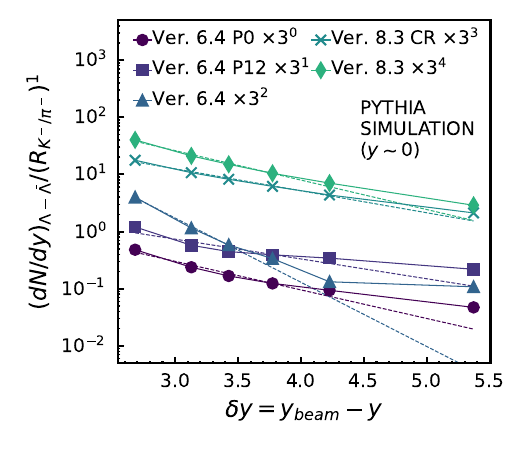}
    \caption{}
    \label{fig:pythiaNetL}
\end{subfigure}
\begin{subfigure}[b]{0.32\linewidth}
    \includegraphics[width=\linewidth]{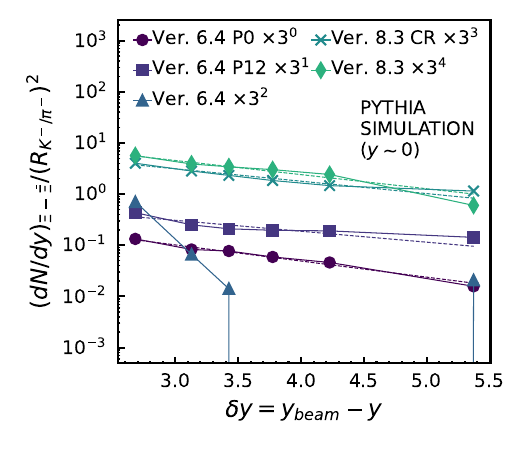}
    \caption{}
    \label{fig:pythiaNetX}
\end{subfigure}
\begin{subfigure}[b]{0.32\linewidth}
    \includegraphics[width=\linewidth]{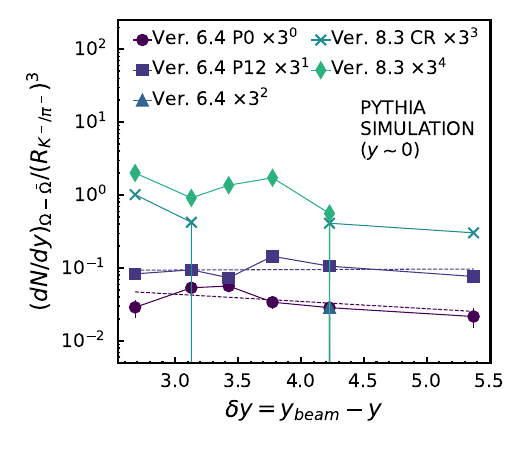}
    \caption{}
    \label{fig:pythiaNetO}
\end{subfigure}
\caption{Predictions of net-$\Lambda$ (a), net-$\Xi$ (b) and net-$\Omega$ (c) yields, scaled by $(R_{K^-/\pi^-})^n$ where $n$ is the number of valence strange quarks in the hyperon, from different versions and tunes of \texttt{PYTHIA}. Only statistical uncertainties are included, which are smaller than the marker size in most cases. Dotted lines represent fits with exponential functions. For certain tunes, some data points are missing due to negative predicted net-hyperon yields, and exponential fits are not performed.}
\label{fig:pythiaNetH}
\end{figure*}

The procedure used for extracting $\alpha_B$ from \texttt{PYTHIA} closely resembles that employed for experimental data. $p$+$p$ collisions at the same center-of-mass energies as those for the Au+Au collisions used in Fig.~\ref{fig:corNet} are simulated. Following the procedure outlined in Refs.~\cite{star:2019sh, STAR:2017sal}, all $K^-$, $\Xi$, $\bar{\Xi}$, $\Omega$, and $\bar{\Omega}$ particles generated by \texttt{PYTHIA} are counted, regardless of whether they come from primary interactions or decay products. For $\pi^-$, those coming from the decay of $K^0_S$, $\Lambda$, $\Xi$ and $\Omega$ are not counted. For $\Lambda(\bar{\Lambda})$, those from $\Xi^-(\bar{\Xi^+})$ and $\Omega^-(\bar{\Omega^+})$ decays are excluded. At the same time, all $\Sigma^0$'s count as $\Lambda$'s. Resulting $R_{K^-/\pi^-}$ within $|y| < 0.1$ are shown in Fig.~\ref{fig:pythiaKpi} for different versions and tunes of \texttt{PYTHIA}. While a similar increasing trend with beam energy is seen as that observed in data, sizable differences exist for different \texttt{PYTHIA} setups. Figure~\ref{fig:pythiaNetH} shows net-$\Lambda$, net-$\Xi$ and net-$\Omega$ yields, scaled by $(R_{K^-/\pi^-})^n$, within a rapidity window of $|y| < 0.5$. Also shown in Fig.~\ref{fig:pythiaNetH} as dotted lines are exponential fits using Eq.~\eqref{eq:exp} to the normalized net-hyperon yields to extract $\alpha_B$ values. In some cases, the exponential function does not describe the $\delta y$ dependence very well. For example, the net-$\Lambda$ yields predicted by \texttt{PYTHIA} are always above the fit function at 200 GeV, and the net-hyperon yields from \texttt{PYTHIA} 6.4 P12 do not align with the exponential shape well. For some \texttt{PYTHIA} tunes, they predict negative net-hyperon yields at certain energies, which are omitted from Fig. \ref{fig:pythiaNetH}. For such cases, no exponential fits are performed.

%$dN/dy_{\Lambda-\bar{\Lambda}}/(dN/dy_{K^-/\pi^-})$ and Fig.~\ref{fig:pythiaKpi} shows $dN/dy_{K^-/\pi^-}$ for different versions of \texttt{PYTHIA}. For a fair comparison, the numerator $dN/dy_{\Lambda-\bar{\Lambda}}$ is calculated with a rapidity window of $|y| < 0.5$ while the denominator $dN/dy_{K^-/\pi^-}$ is calculated with a rapidity window of $|y| < 0.1$. Only statistical uncertainty is plotted for each individual data point. 

%Although \texttt{PYTHIA} output indicates which particle is primary and which one is decay products, it is still advisable not to use this information to select primary yields. We choose to perform feed-down correction on \texttt{PYTHIA} output in a way that is consistent with feed-down correction being performed on data. Following the procedure from Refs.~\cite{star:2019sh, STAR:2017sal}, all $K^-$, $\Xi$, $\bar{\Xi}$, $\Omega$, and $\bar{\Omega}$ particles generated by \texttt{PYTHIA} are counted, regardless of whether they come from primary interactions or decay products. For $\pi^-$, only those which come from the decay of $K^0_S$ are not counted. For $\Lambda(\bar{\Lambda})$, only those which come from the decay of $\Xi^-(\bar{\Xi^+})$ and $\Omega^-(\bar{\Omega^+})$ are not counted. At the same time, all $\Sigma^0$ counts as $\Lambda$, in accordance with the way data is analyzed~\cite{star:2019sh}.

%Furthermore, some models predict an increase in net-hyperon yields with $\delta y$, which contradicts experimental measurements. If the fitted $\alpha_B$ is negative, the fit is considered to have failed and these results are omitted as well.

\begin{comment}
Systematic uncertainties in the extracted $\alpha_B$ values are estimated by fitting net-hyperon yields over different ranges of $\delta y$. The central values are obtained with the fitting range of $0 < \delta y < 6$, while $0 < \delta y < 4.3$ and $0 < \delta y < 4$ are used as systematic variations. The resulting systematic uncertainty is calculated as the following:
\begin{equation}
\sigma_{\text{sys}} = \sqrt{\frac{(\alpha_{B,\delta y < 6} - \alpha_{B,\delta y < 4.3})^2 + (\alpha_{B,\delta y < 6} - \alpha_{B,\delta y < 4})^2}{2}}.
\end{equation}
\end{comment}

 Table~\ref{tab:ExpSlope} tabulates predicted $\alpha_B$ values for net-hyperons from various versions and tunes of \texttt{PYTHIA}. They are compared to experimentally measured values in the 0-80\% centrality class except for net-$\Omega$ which is from 0-60\% centrality class. Results for net-$p$ from Ref.~\cite{lewis2023search} are also included. 
 %Errors listed in the table are total uncertainties, which are quadratic sums of statistical and systematic uncertainties, for both data and \texttt{PYTHIA}. 
 Values for some configurations of \texttt{PYTHIA} are marked as ``N.A.", corresponding to cases where the \texttt{PYTHIA} distributions were not fitted due to the presence of negative yields at certain beam energies.

% I think we can keep the favor dependence paragraph with this updated uncertainty
All \texttt{PYTHIA} configurations show strong flavor dependence of $\alpha_B$. Their predicted $\alpha_B$ values for net-$\Lambda$ are collectively larger than those for net-$\Xi$, while a further decrease in $\alpha_B$ is seen for net-$\Omega$. However, these trends are not observed in data. Compared to data, \texttt{PYTHIA} predictions generally overshoot the value of $\alpha_B$ for net-$\Lambda$ and underestimate it for net-$\Xi$. For net-$\Omega$, only two \texttt{PYTHIA} 6.4 tunes can produce meaningful slopes, both of which are still significantly smaller than that observed in data. Considering net-$p$, \texttt{PYTHIA} 6.4 P12 predicts a slope that is a factor of 1.7 smaller than the measured value, while the predicted value from \texttt{PYTHIA} 8.3 is about 1.6 times larger.

\texttt{PYTHIA} 8.3 CR performs better than its non-CR counterpart as its predicted $\alpha_B$ values are closer to data for both net-$p$ and net-$\Lambda$, which underscores the potential of incorporating, even partially, the baryon junction mechanism. However, it still fails to reproduce the slope parameters for net-$\Xi$ and net-$\Omega$, and does not eliminate the flavor dependence as the $\alpha_B$ values change significantly between net-$\Lambda$ and net-$\Xi$. This indicates the need for a genuine inclusion of the baryon junction mechanism in event generators to be tested against data.

\begin{comment}
\begin{figure}[h]
\centering
\begin{subfigure}[b]{0.64\linewidth}
    \includegraphics[width=\linewidth]{Pythia.pdf}
    \caption{}
    \label{fig:ExpSlope}
\end{subfigure}

\caption{(a): $\alpha_B$ as a function of centralities measured by BES-I program. The $x-$errorbar on $\Omega$ indicates it's widen centrality bins as it spans 0-60\%. (b): $\alpha_B$ when particle yields in 0-60\% centrality is summed and that predicted by various \texttt{PYTHIA} versions. Each row corresponods to a different particle species. There are arrows pointed towards the upper and lower edge of each row, indicating off-scale high and off-scale low value of $\alpha_B$, respectively.}
\end{figure}
\end{comment}

\begin{table*}[t]
    \centering
        \caption{$\alpha_B$ values for different baryon species extracted from 0-80\% Au+Au collisions, except for $\Omega$ which uses 0-60\% centrality due to lack of peripheral data. $R_{K^-/\pi^-}$ from both 0-10\% central Au+Au collisions and $p$+$p$ collisions are used to account for strangeness production suppression. The listed experimental errors represent combined statistical and systematic uncertainties. $\alpha_B$ values predicted by various versions and tunes of \texttt{PYTHIA} are also listed for comparison, and N.A. appears when the \texttt{PYTHIA} configuration predicts negative net-hyperon yield at any beam energy and no exponential fit is performed.}
    \label{tab:ExpSlope}
    \begin{tabular}{c|cc|ccccc}
         & \multicolumn{2}{c|}{Au+Au (0-80\%)} & \multicolumn{5}{c}{\texttt{PYTHIA}}\\
        \hline
        Species & Au+Au $R_{K^-/\pi^-}$  & $p$+$p$ $R_{K^-/\pi^-}$& Ver. 6.4 & Ver. 6.4 (P0) & Ver. 6.4 (P12) & Ver. 8.3 & Ver. 8.3 CR Mode 2\\
        \hline
        $p$ & $0.64 \pm 0.05$ & - & $0.86 \pm 0.05$ & $0.76 \pm 0.03$ & $0.38 \pm 0.02$ & $1.01 \pm 0.03$ & $0.73 \pm 0.02$ \\
        $\Lambda$ & $0.72 \pm 0.06$ & $0.77 \pm 0.06$ & $2.58 \pm 0.03$ & $1.15 \pm 0.01$ & $0.80 \pm 0.01$ & $1.19 \pm 0.01$ & $0.89 \pm 0.01$ \\
        $\Xi$ & $0.86 \pm 0.10$ & $0.95\pm0.11$ & N.A. & $0.73 \pm 0.05$ & $0.49 \pm 0.05$ & $0.64 \pm 0.08$ & $0.56 \pm 0.06$ \\
        $\Omega$ & $0.97 \pm 0.28$ & $1.09\pm 0.28$ & N.A. & $0.23 \pm 0.10$ & $-0.01 \pm 0.15$ & N.A. & N.A. \\
    \end{tabular}

\end{table*}

\section{Summary}
Continuing from the analysis of baryon transport, approximated using inclusive net-$p$, in Ref.~\cite{lewis2023search}, we investigate the validity of the baryon junction picture by analyzing net-hyperon yields in Au+Au collisions from RHIC BES-I program. After accounting for the effect related to the difficulty associated with the strange quark production by dividing net-hyperon yields with $(R_{K^-/\pi^-})^n$, where $R_{K^-/\pi^-}$ is the $K^-/\pi^-$ yield ratio and $n$ is the number of valence strange quarks in a hyperon, net-hyperon yields at mid-rapidity ($|y| < 0.5$) exhibit the expected exponential dependence on beam rapidity. The extracted exponential slopes ($\alpha_B$) for net-$\Lambda$, net-$\Xi$, and net-$\Omega$ are consistent with each other and with those of net-$p$, which suggests a common transport mechanism for different hyperons and 
%The value of $\alpha_B$ for protons is smaller than those of hyperons, possibly due to $K^-/\pi^-$ not being a perfect gauge for strangeness production suppression. Nonetheless, he $\alpha_B$ values are still close to each other, which 
supports a flavor-blind baryon junction picture. 
Predictions from different versions and tunes of \texttt{PYTHIA} event generator, which do not include baryon junctions in the colliding protons or involve junctions in scatterings, are compared to data. While some tunes can describe the net-$p$ slope, they are unable to consistently replicate the slope parameters for both net-$p$ and net-hyperons. With significantly enhanced statistics from RHIC BES-II program, hyperon yields at mid-rapidity can be measured with much better precision than BES-I, which will provide more rigorous tests for baryon transport dynamics. 

\section*{Acknowledgments}

The authors would like to thank Flemming Videbaek, Dimitri Kharzeev, Huan Z. Huang, Xianglei Zhu and Lokesh Kumar for valuable discussions. This research is supported by the US Department of Energy, Office of Nuclear Physics (DOE NP), under contract Nos. DE-FG02-89ER40531, DE-SC0012704, DE-FG02-10ER41666, and DE-AC02-98CH108.

\section*{Appendix}
\subsection*{$K^{-}/\pi^{-}$ yield ratio in $p$+$p$ collisions}
\label{sect:app-kpi-ratio}
\begin{figure}[h!]
\centering
    \includegraphics[width=0.85\linewidth]{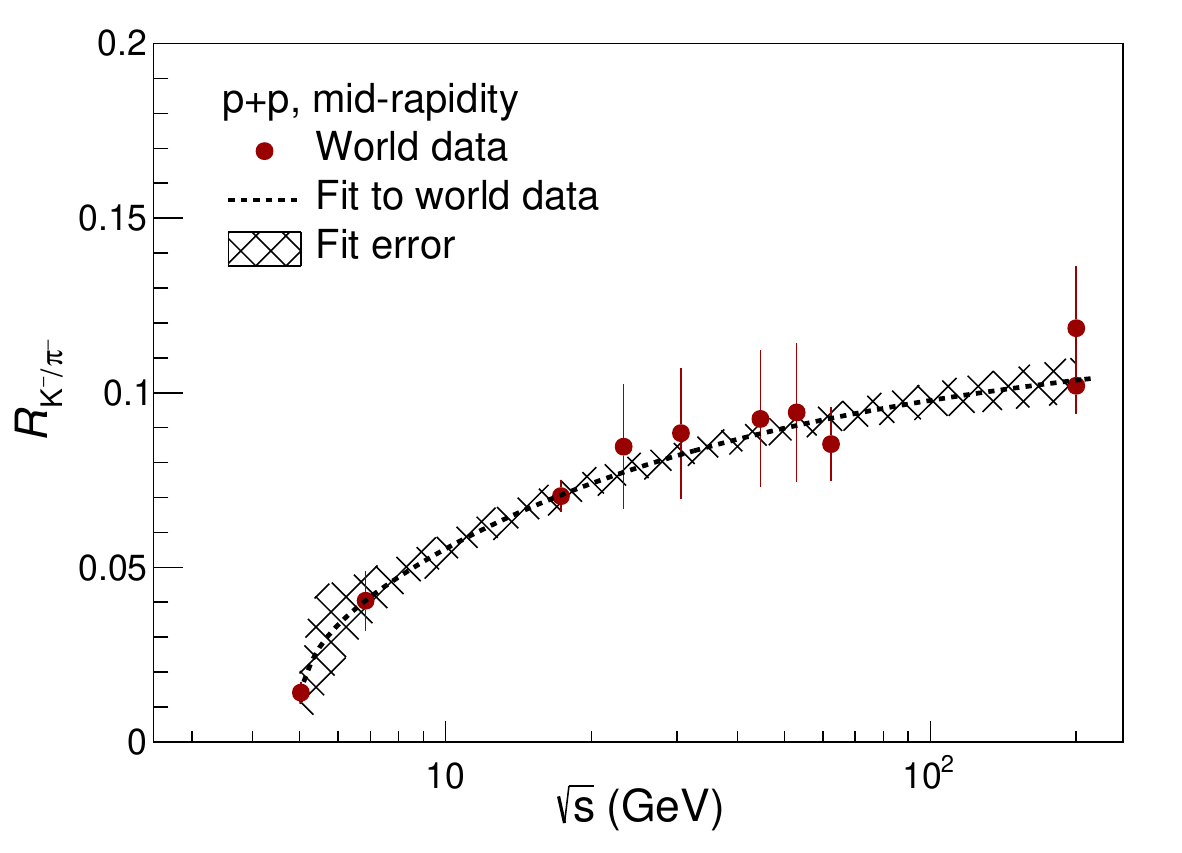}
    \caption{Parametrization of world data on $R_{K^-/\pi^-}$ at mid-rapidity as a function of collision energy in $p$+$p$ collisions.}
    \label{fig:kpi-pp-fit}
\end{figure}

Mid-rapidity $K^{-}/\pi^{-}$ yield ratios ($R_{K^-/\pi^-}$) in $p$+$p$ collisions at center-of-mass energies ($\sqrt{s}$) of 7.7, 11.5, 19.6, 27, 39, 62.4 and $\SI{200}{GeV}$ are obtained from parametrization of world data on $R_{K^-/\pi^-}$ as a function of collision energy. Multiplicities of $K^{-}$ and $\pi^{-}$ within the full rapidity range are measured at ISR energies of 23.4, 30.6, 44.6 and 52.8 GeV~\cite{Rossi:1974if} as well as at 5.0 and 6.8 GeV~\cite{Antinucci:1972ib}. A total uncertainty of 15\% is assumed for both the $K^{-}$ and $\pi^{-}$ yields~\cite{Rossi:1974if,Antinucci:1972ib}. To convert $R_{K^-/\pi^-}$ from full rapidity to mid-rapidity, a factor of 1.323 is applied based on the rapidity distributions of $K^{-}$ and $\pi^{-}$ measured at ISR energies~\cite{Antinucci:1972ib}. In addition, yields of $K^{-}$ and $\pi^{-}$ at mid-rapidity are measured in $p$+$p$ collisions at $\sqrt{s}$ = 17.3 GeV by the NA49 experiment at SPS~\cite{NA49:2005qor,NA49:2009wth} and at $\sqrt{s}$ = 62.4 and 200 GeV by the PHENIX experiment at RHIC~\cite{PHENIX:2011rvu}. For the NA49 results, the uncertainty upper limits of 4.5\% and 4.8\% are used for $K^{-}$ and $\pi^{-}$ measurements respectively. For all these measurements, it is assumed that uncertainties in $K^{-}$ and $\pi^{-}$ yields are uncorrelated when calculating $R_{K^-/\pi^-}$. In addition, the STAR experiment at RHIC reported $R_{K^-/\pi^-}$ at mid-rapidity in 200 GeV $p$+$p$ collisions \cite{STAR:2008med}. The collection of world data is shown in Fig. \ref{fig:kpi-pp-fit}. They are fitted with the function~\cite{Antinucci:1972ib}: 
\begin{equation}
f(\sqrt{s})=\frac{A+2Bln(\sqrt{s})+C/\sqrt{s}}{A'+2B'ln(\sqrt{s})+C'/\sqrt{s}},
\end{equation}
where $A,B, C, A', B', C'$ are free parameters. 
The resulting fit function is shown as the dashed curve in Fig. \ref{fig:kpi-pp-fit} along with its error band. The mid-rapidity $R_{K^-/\pi^-}$ at desired energies are obtained from the fit function and listed in Table \ref{tab:kpi-pp-exp}.
%In addition to the fit error, two more sources of uncertainties are considered: i) a factor of 1.276 to convert $K^{-}/\pi^{-}$ ratios at full rapidity to mid-rapidity is estimated based on NA49 results~\cite{NA49:2005qor,NA49:2009wth} and applied; ii) a third-order polynomial function against $ln(\sqrt{s})$ is used to fit the collision energy dependence of $K^{-}/\pi^{-}$ ratio. The total uncertainties are estimated by summing up the three individual sources in quadrature and listed in Table \ref{tab:kpi-pp-exp} as well.
\begin{table}[hbt]
    \centering
        \caption{Extrapolated $R_{K^-/\pi^-}$ at mid-rapidity, along with their uncertainties, in $p$+$p$ collisions at various energies.}
    \label{tab:kpi-pp-exp}
    \begin{tabular}{c|c}
    \hline
    $\sqrt{s}$ [GeV] & $dN/dy_{K^{-}/\pi^{-}}$ ($y\sim0$)\\
    \hline
    7.7 & $0.0457\pm0.0065$\\
    11.5 & $0.0597\pm0.0046$\\
    19.6 & $0.0736\pm0.0042$\\
    27 & $0.0801\pm0.0039$ \\
    39 & $0.0863\pm0.0038$ \\
    62.4 & $0.0927\pm0.0042$ \\
    200 & $0.1036\pm0.0066$ \\
    \hline
    \end{tabular}
\end{table}

\subsection*{Net-proton rapidity slope parameter in $p$+$p$ collisions}
\label{sect:app-netp-slope}
Rapidity slope parameter ($\alpha_B$) for inclusive net-$p$ in $p$+$p$ collisions is extracted by fitting its yield as a function of $\delta y = y_\text{beam} - y$ with an exponential function: 
\begin{equation}
f(\delta y) = C\times e^{-\alpha_B\delta y}
\end{equation}

\begin{figure}[h!]
\centering
    \includegraphics[width=0.85\linewidth]{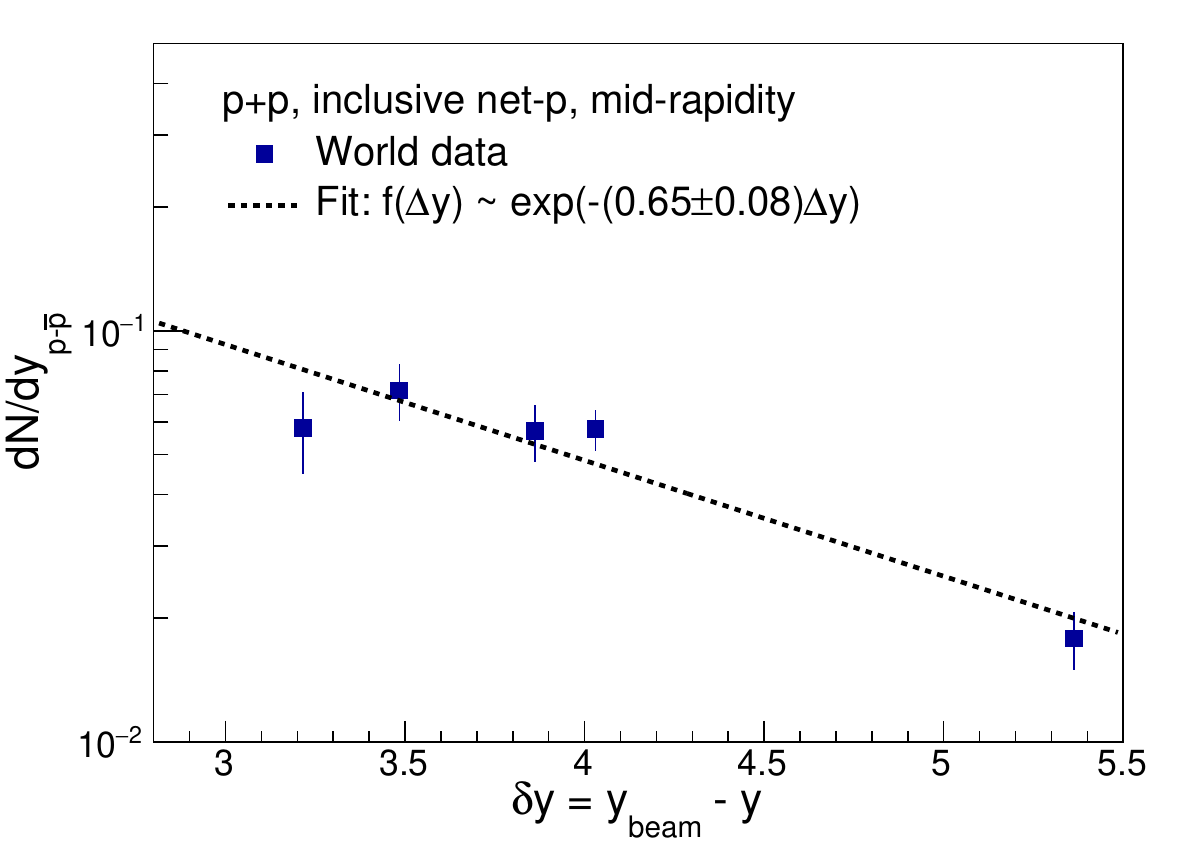}
    \caption{Fit to inclusive net-$p$ yield at mid-rapidity as a function of $\delta y$ in $p$+$p$ collisions.}
    \label{fig:netp-pp-incl-fit}
\end{figure}

Inclusive proton and anti-proton yields, which include both the primordial production and those from weak decays, at mid-rapidity are measured in $p$+$p$ collisions by experiments at the ISR~\cite{Banner:1972eja,British-Scandinavian:1974lzd,Antinucci:1972ib} and the STAR experiment at RHIC~\cite{STAR:2008med}. For ISR measurements, differential cross sections as a function of transverse momentum (\pt) are fit with the function $f(p_{\rm T}) = A\times e^{Bp_{\rm T}+Cp_{\rm T}^{2}}$ to obtain the total cross sections. They are then converted to rapidity densities ($dN/dy$) using inelastic $p$+$p$ cross sections based on its parametrization as a function of beam energy~\cite{STAR:2020phn}. Since only statistical errors are included in the fitting, errors on proton and anti-proton $dN/dy$ are treated as uncorrelated when calculating net-$p$ yields. In addition, luminosity uncertainties (6\% for $\sqrt{s}$ = 44.6 and 52.8 GeV, 10\% for 30.6 GeV and 15\% for 23.4 GeV)~\cite{British-Scandinavian:1974lzd} are included in the net-$p$ yield uncertainties. STAR measures net-$p$ yield in 200 GeV non-singly diffractive $p$+$p$ collisions. To covert to inelastic collisions, a scale factor of $\sigma_{\rm inel}/\sigma_{\rm NSD} = (30\ \rm mb)/(42\ \rm mb) = 0.714$ is multiplied to the STAR result~\cite{STAR:2008med}. The resulting inclusive net-$p$ yields at mid-rapidity are summarized in Table \ref{tab:netp-pp-incl} for $p$+$p$ collisions of various energies, and displayed in Fig.~\ref{fig:netp-pp-incl-fit} as a function of $\delta y$. An exponential fit to the $\delta y$ dependence yields $\alpha_{B} = 0.65\pm0.08$. 
\begin{table}[hbt]
    \centering
        \caption{Inclusive net-$p$ yields, including both the primordial production and those from weak decays, at mid-rapidity in $p$+$p$ collisions at various energies.}
    \label{tab:netp-pp-incl}
    \begin{tabular}{c|c}
    \hline
    $\sqrt{s}$ [GeV] & $dN/dy_{\mathrm{net}-p}$ ($y\sim0$) \\
    \hline
    23.4 & $0.058\pm0.013$\\
    30.6 & $0.072\pm0.011$\\
    44.6 & $0.057\pm0.009$\\
    52.8 & $0.058\pm0.007$ \\
    200 & $0.018\pm0.003$~\cite{STAR:2008med} \\
    \hline
    \end{tabular}
\end{table}

\biboptions{numbers,sort&compress}
\bibliographystyle{elsarticle-num} 
\bibliography{reference}

\end{document}